\documentstyle[prl,aps,epsfig,floats]{revtex}
\begin{document}
\twocolumn[
\hsize\textwidth\columnwidth\hsize\csname@twocolumnfalse\endcsname
\draft
\title{{Electronic Mechanism of Superconductivity in
the Cuprates, C$_{60}$, and Polyacenes}}

\author{Sudip Chakravarty and Steven A. Kivelson}
\address{Department of Physics and Astronomy, University of California Los
Angeles
\\ Los Angeles, California 90095-1547}
\date{\today}
\maketitle
\begin{abstract}
On the basis of an analysis of
theoretical and numerical studies of model systems,
and of experiments on
superconductivity in doped C$_{60}$,  polyacenes,
and the cuprate high temperature superconductors, we propose
that a purely electronic mechanism of
superconductivity requires structures at an
intermediate or  mesoscale.  Specifically,
we address the crucial question of how  high temperature superconducting
pairing on the mesoscale can arise from
purely repulsive electronic interactions.

\end{abstract}

\pacs{PACS:}
]
\section{Introduction}

Induced attractions between electrons, such as are mediated by phonons
in  conventional
superconductors, are typically weak while the direct repulsive
interactions between electrons are strong.  It has therefore long been
felt that the key to high temperature superconductivity is a purely
electronic mechanism that  directly exploits the repulsive
interaction to produce pairing.

After the discovery of high temperature superconductivity in the
cuprates by Bednorz and M{\"u}ller\cite{bm}, an idea for how this could be
achieved was proposed\cite{pwa}.  It was suggested that an
antiferromagnet disordered by quantum fluctuations and characterized by
resonating singlet
pairs of
spins  could be viewed as a state  in which superconducting pairing occurs,
even
though the
system
is an insulator with one electron per site and
zero superfluid density!  In this case, when doped,
so that charge motion is allowed,
the system will automatically become a superconductor.
A related idea is that the singlet pair formation in
the insulating state produces a
gap in the spin excitation spectra\cite{krs,rbl}, which evolves  into the
superconducting
pairing gap upon doping.

A fundamental flaw with this approach soon became clear;  in the
relevant two-dimensional spin 1/2 antiferromagnet on a square
lattice, the ground state is magnetically ordered\cite{chn}, and
indeed so strongly  that the elementary excitations about the ordered state,
spin waves, give
a quantitatively accurate description of many quantities.
 Needless to say,  the
magnetically ordered state bears no resemblance to a
superconductor, even locally!

Yet, the intuition that the high energy scale of high temperature
superconductors
must
have its origin in the repulsive interactions remained. The
discovery by Hebard {\em et al.}\cite{hebard}
of high
temperature superconductivity  in alkali doped  C$_{60}$ was another
milestone.
It illuminated a path, in the clearest terms,
that the missing element in the above revolutionary argument is the
existence of a mesoscale
structure. The C$_{60}$ molecule itself plays the role of this structure
permitting
electronic pairing, which  would disappear if the molecule
were
expanded to the macroscopic scale. We
were thus led to propose a mechanism\cite{ck,cgk,baskaran} of superconductivity
in which attraction arises from repulsion.
Later it was
suggested that
a similar pairing mechanism may apply in the cuprate high temperature
superconductors\cite{ek,ekz}, where the relevant mesoscopic
structures are self-organized stripes.

Here, we present evidence, which we consider to be exceedingly strong,
that this set of ideas underlies the
occurrence of high temperature superconductivity in diverse materials.
It is shown that local singlet
formation as a driving mechanism for
``pairing'' is correct, but only on a mesoscopic scale.  The pairing
tendencies
are either much
weaker or non-existent in extended systems.

There are also  a number of
remarkable recent experiments on  doped C$_{60}$\cite{bertram1,bertram2}
and polyacene crystalline films\cite{bertram3}
which encourage the view  that these are electronically driven true high
temperature superconductors in disguise. Although the authors of these
papers
ascribe the superconductivity of these materials  to the
electron-phonon interaction, we feel that the experimental facts,
including the high value of $T_{c}$ itself (presently up to 52K),
rather indicate that these
materials are true ``high temperature superconductors,'' in which
correlation effects play a dominant role both in the normal state and
in the mechanism of superconductivity.  We shall briefly discuss this
issue in the final section.

It is important to stress the distinction between {\it possible} and
{\it necessary} consequences of the existence of mesoscale
structures --- surely not all forms of
repulsive interactions on any given cluster will lead to pair binding.
Additionally, there are strong
reasons\cite{kfe,cnl} to expect the same mesoscale physics to lead
to  competing  orders.  Indeed, we propose that
the existence of a complex phase diagram with a variety of
broken symmetry phases, including high temperature superconductivity,
is a characteristic feature of a material with an electronic
mechanism of pairing on the mesoscale.

\section{Mesoscale pair-binding}

 At atomic
scales, the Hamiltonian is simple (although strongly
interacting), and the dominant Coulomb interactions are readily identified.
At the macroscopic scale, the physics is again simple and is governed by the
universal properties of the attractive fixed points of the renormalization group
theory.
However, as the
high energy degrees of freedom are integrated out, beginning with the
microscopic scale, every
imaginable multiparticle interaction is generated, so that  the effective
Hamiltonian becomes
complicated and the dominant physics can  be obscure.
Thus,
it is necessary to first
determine whether
there
is any robust and  predictable behavior at all.
We argue that on certain finite clusters quantum fluctuations of the spins
can
produce a disordered
ground-state with a large gap in the spin excitation spectrum
and a robust
pair-binding of holes or electrons.

In each case, we start with a ``neutral'' cluster with $N$ sites and one
electron per site.
For all the systems studied, the neutral cluster has a unique, spin
singlet groundstate.
We define $E(Q,S)$ to be the lowest energy eigenvalue of the cluster
with total spin, $S$, and
``charge,''$Q$, or equivalently a total of
$N+Q$ electrons.
Whenever the ground-state is a spin singlet, we can define the
spin gap $\Delta_{s}$ to be
\begin{equation}
\Delta_{s}(Q)=E(Q,1)-E(Q,0).
\end{equation}
The pair-binding energy is defined as
\begin{equation}
E_{p}(Q)=2E(Q+1)-E(Q+2)-E(Q)
\end{equation}
where $E(Q)$ signifies the minimum over $S$ of $E(Q,S)$.  Clearly, a
positive pair-binding energy signifies an effective attraction between
electrons, in the sense that given $2(N+Q+1)$ electrons and two
clusters, it is energetically preferable to place $N+Q+2$ electrons
on one cluster and $N+Q$ on the other than to put $N+Q+1$ electrons on
each of the clusters.

We illustrate the origin of these effects, the relation
between  $\Delta_s$ and $E_p$, and the importance of
intermediate scales to the occurrence of such ``attraction from
repulsion''
by analyzing a variety of  models.
We defer to the following section any
discussion of the relation
between spin-gap formation and pair-binding with the occurrence of
superconducting or other long-range order in extended systems.

\subsection{Hubbard rings}

As a first example consider Hubbard rings with $N$ sites, where $N$ is
even. The
Hamiltonian is
\begin{equation}
 H=-t\sum_{i\sigma} (c_{i\sigma}^{\dagger}
c_{i+1\sigma}+ h.c.)
 +\frac U 2 \sum_{i\sigma}n_{i\sigma}n_{i-\sigma}
\end{equation}
with the implied periodic boundary condition.
The fermion operator $c^{\dagger}_{i\sigma}$ creates an
electron with spin $\sigma$ on site $i$, and
$n_{i\sigma}=c^{\dagger}_{i\sigma} c_{i\sigma}$ is the density of electrons
of
spin $\sigma$ on site $i$.

\begin{figure}[htb]
\centerline{\epsfxsize=\linewidth\epsffile{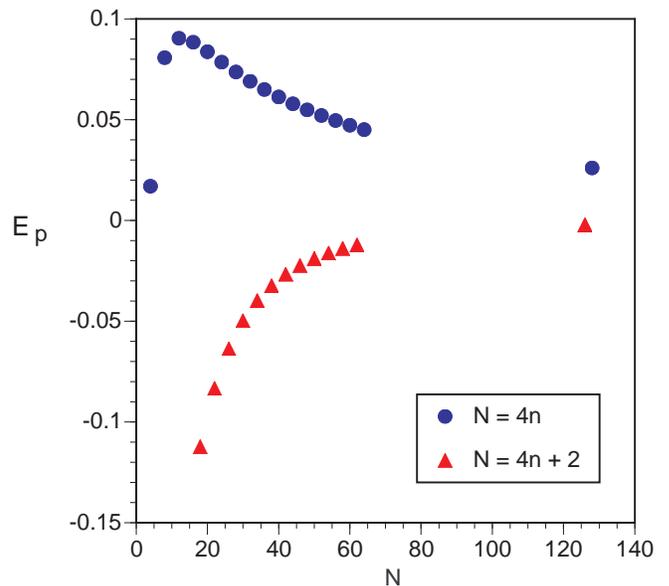}}
\caption{Pair-binding energy, $E_p$, of $N=4n$ and $N=4n+2$ site Hubbard
rings
with $t=1$ and $U=4$.}
\label{fig1}
\end{figure}

The energies, $E(Q,S)$, can be calculated exactly from
Bethe Ansatz\cite{fye}. We have extended existing results for the
pair-binding energy, $E_p$, by
numerically evaluating the Bethe-Ansatz equations for a wide range of
system sizes as large as 1024 sites, and by calculating 
the spin-gaps,
$\Delta_s$. The exact particle-hole symmetry of the Hubbard
model on a bipartite lattice implies that electron doping, $Q>0$,
and hole doping, $Q<0$, are equivalent. As can be seen in
Fig.~\ref{fig1}
pair-binding occurs for electrons added to the neutral molecule,
$E_{p}(0)>0$, whenever $N=4n$, but does not occur when $N=4n+2$,
where $n$ is a positive integer.
The difference between these two cases
can be readily understood
from low order perturbation theory in $U/t$ as discussed in Appendix~\ref{sec:append}.

\begin{figure}[htb]
\centerline{\epsfxsize=\linewidth\epsffile{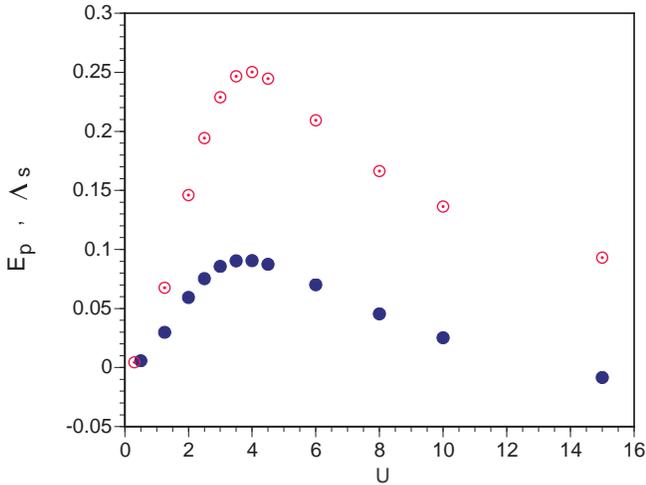}}
\caption{Pair-binding energy, $E_{p}$ (solid symbol), and spin-gap,
$\Delta_{s}$
(hollow symbol), of
12 site Hubbard ring
as a function of $U$ in units of $t=1$.}
\label{fig2}
\end{figure}

The role of intermediate scales  can be
seen directly from Fig.~\ref{fig1};  the pair-binding energy
vanishes for large $N$ and is maximum at an intermediate value of $N$.
Moreover, in Fig.~\ref{fig2} we show $E_p$ for fixed $N$ as a
function of $U$, from which it is clear that it is maximal for
intermediate interaction strengths as well.  Indeed,
we have proven
a rigorous theorem\cite{cck} that for the standard Hubbard model on any
finite
lattice and with
positive hopping matrix elements, $E_{p}\le 0$ for $U\to\infty$.
(Trivially, $E_{p}\le 0$ for $U= 0$.)

\begin{figure}[htb]
\centerline{\epsfxsize=\linewidth\epsffile{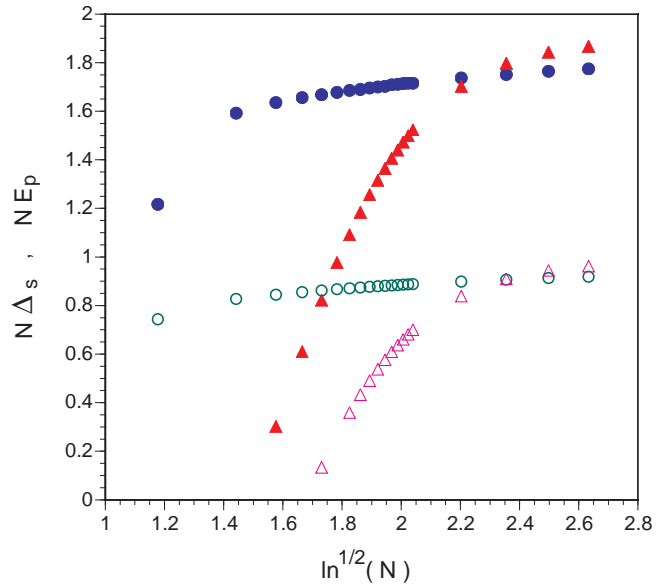}}
\caption{The large $N=4n$ scaling of the pair-binding energy,
$E_{p}$ (triangles), and spin-gap, $\Delta_{s}$ (circles)
of Hubbard rings in units of $t=1$.  All energies are scaled by an explicit
factor of $N$ to remove the dominant $1/N$ dependence, revealing the
logarithmic dependence predicted in Eq. (\ref{eq:ds}).  Solid
figures are for $U=10$ and open figures are for $U=20$.}
\label{fig3}
\end{figure}

The crucial question is what is the mechanism of pair-binding? We can gain
some
insight into this
question in the large-$N$ limit from bosonization\cite{bosonization}.  We show that
pair-binding is
closely related to  the phenomenon of
spin-gap formation. It is well-known that at  long wavelength, spin-charge
separation occurs in
1D Hubbard rings.  Therefore, all energies can be written as a sum of  a
spin
contribution and a charge
contribution. The result is that for $N=4n\gg 1$,
\begin{eqnarray}
\Delta_s&=&\frac{v_s}{N}\left[B_1\ln^{1/2}(N)+B_2\right] +
\ldots\label{eq:ds}\\
E_p&=&\Delta_s +B_3 \frac {v_s} N- \frac{B_4}{N^2}
\left[\frac{{v_c}^2}{\Delta_c}\right]+
\ldots\label{eq:epair}
\end{eqnarray}
Here, $v_{s}$ and $v_{c}$ are the spin and charge velocities,
respectively (in units in which the lattice constant is unity) and
$\Delta_{c}$ is the charge-gap in the $N\to \infty$ limit.
The constants, $B_j$, are numbers of order unity, which we have
not evaluated.  The large-$N$ behavior of these quantities computed
from the Bethe Ansatz  are shown in Fig.~\ref{fig3} and appears to be in
good agreement with these expressions.

Because of the
relevance of Umklapp scattering,
%($N=4n$ sequence),
the charge degrees of freedom
are
described by a sine-Gordon field theory.  Dilute doping results in a dilute
gas
of spin 0 charge
$\pm e$ solitons
(holons or eons\cite{cgk}) with a gapped spectrum.  These  solitons can,
in turn, be mapped onto massive Dirac fermions.
Consequently, the finite size level spacings are those of a
non-relativistic particle in a box, which gives rise to the third
term in Eq.~\ref{eq:epair}.  However, in making a spin excitation
in the neutral molecule, no charge is added, so the spin-gap is
independent of the charge dynamics.

It has been previously observed\cite{woynarovic,korepin} that, because the
spin-system is quantum critical, the spin correlation functions are
conformally invariant, so that the spin-gap must scale with
$v_{s}/N$.  The slow renormalization group
flow associated with a marginally irrelevant operator at the fixed
point invalidates this finite size scaling behavior, however.  A
consequence of this for the infinite system is that the  spin-spin
correlation
function falls with
the distance, $r$, as
$(\ln^{1/2}r)/r$\cite{Affleck}.  We conjecture that a similar logarithmic
correction occurs in the finite size scaling behavior, hence the
logarithmic term in Eq. (\ref{eq:ds}).  As can be seen from the large
$N$ behavior of the Bethe Ansatz solution in Fig.~\ref{fig3}, just such a
logarithmic term appears
in the data.  In terms of the elementary spinon excitations of the spin 1/2
Heisenberg model, $\Delta_s$ is interpreted as the energy of a two-spinon
state,
while $\Delta_s+B_3v_s/N$ is twice the single spinon energy;  $-B_3v_s/N$
thus
has
the interpretation of an effective spinon-spinon interaction.

An important lesson of
this analysis is that, for large $N$, $E_{p}\approx \Delta_{s}$.
This relation, originally proposed by us in the context of
pair-binding on a $C_{60}$ molecule\cite{cgk}, is in this case a consequence
of
the quantum critical character of the 1D Hubbard model.  Moreover,
since both the separation of charge and spin, and the quantum critical
nature of the spin correlations, are robust features of the one
dimensional electron gas, they are not limited to the integrable
Hubbard model, but would be expected to survive the inclusion of
second and further neighbor Coulomb repulsions as well.

\subsection{$t-J$ ladders}

Hubbard or $t-J$ chains have no spin-gap in the limit $N\rightarrow \infty$,
and, in addition, have  small superconducting susceptibilities irrespective
of
the  doping
level. In contrast, ladder systems, that is,  quasi-one-dimensional systems
of
finite width, can
exhibit both spin-gap  and strong tendencies toward superconducting order,
even
in this
limit.  While these systems are infinite in extent, the mesoscopic physics
comes
in
through the finiteness of the transverse dimension.  Because so much is
known
theoretically about one-dimensional quantum systems, such ladders are also
an
excellent
laboratory for exploring the general physical principals set forth in the
Introduction.

\begin{figure}[htb]
\centerline{\epsfxsize=3 in\epsffile{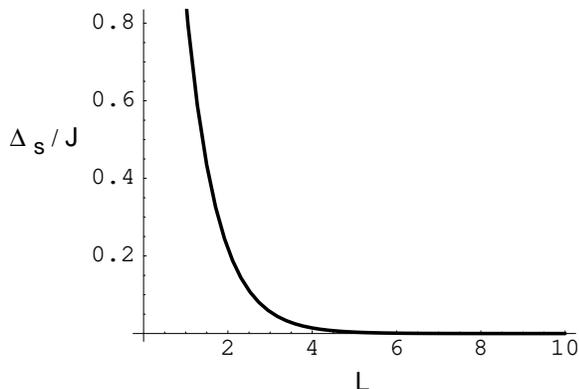}}
\caption{Semi-infinite ladder of coupled quantum $S=1/2$ Heisenberg spins;
the
width of the
ladder is $2L$. The plot shows that the spin gap $\Delta_s/J$ as
a
function of the width of the ladder falls off exponentially fast.}
\label{fig4}
\end{figure}

The Hamiltonian of a  spin-1/2 Heisenberg  ladder, as shown in
Fig.~\ref{fig4},
is
\begin{equation}
H_J=J \sum_{<ij>}[{\bf S}_i\cdot{\bf S}_j -(1/4)n_in_j]
\end{equation}
where ${\bf S}_i$ is a spin 1/2 operator,
$J$ is the antiferromagnetic exchange interaction, and $\langle ij\rangle$ signifies
nearest-neighbor sites, of spacing $a$, on the ladder. For a Heisenberg
ladder,
the site occupation
numbers,
$n_i$, are  constants of motion and are unity. So, the second term in the
Hamiltonian is simply an
irrelevant constant; this term will become important for the $t-J$-ladder to
be
discussed later. For
an even-leg ladder of width  $W/a=2L=2$,  the spin-gap is known
numerically\cite{2leg} to be
$\Delta_{s}\approx J/2$, but
this gap vanishes exponentially\cite{sudip}, as $\Delta_{s}\sim 3.35
J\exp{[-0.682 (W/a)]}$, for wider ladders, $L=2,3, \ldots$, as shown in
Fig.~\ref{fig4}.
In contrast, for odd-leg ladders, $W/a=2L+1$, $H_J$ has gapless
spinon excitations, but here, too, there is an important energy scale, with
the
same functional
dependence on $L$ as $\Delta_s$, below which the physics is dominated by
spinons, as
for a spin $1/2$ Heisenberg chain.  Since, once again, the spin-gap
formation is
related to superconductivity, as we shall show below, this implies that only
rather narrow ladders (2
or 3-leg) are good candidates for the mesocopic building blocks of a high
temperature superconductor.

For a doped system, we can not only ask about  the dependence $\Delta_s(x)$
of the spin-gap and of the pair binding energy $E_p(x)$
on  the concentration of added charge, $x$, but we can also ask about the
superconducting susceptibility.  To be concrete, we consider results for the
 $t-J$  model
\begin{equation}
H_{tJ}=
-t\sum_{<i,j>,\sigma}[c^{\dagger}_{i,\sigma}c^{\dagger}_{j,\sigma}+{\rm
h.c.}]
+H_J ,
\end{equation}
which  is defined on the space of no doubly occupied sites, and can be
viewed as
the
large $U$ limit of the Hubbard model so long as $J=4t^2/U\ll t$.

Numerical studies\cite{compare} on the two leg $t-J$ ladder with $J=0.35 t$
show
that the
spin-gap is roughly the same magnitude as in the undoped two-leg ladder for
a
range of
$x$ between 0 and  0.15 and so is the pair-binding energy.  The
three-leg ladder\cite{3leg} is more interesting --- the spin gap is 0 for $x
\le
0.05$,
then rises to a value comparable to the spin-gap in the two-leg Heisenberg
ladder for
$x\sim 0.15$, and then becomes small or possibly zero as $x$ gets  to be 0.2
or
larger.
This behavior can be readily understood (and was to a
large extent anticipated)  from both perturbative renormalization group
analyses\cite{balents,zachar} and from strong coupling
bosonization\cite{ekz,zachar} methods.

Because continuous symmetries cannot be broken in one dimension, the ladders
can
never
be truly superconducting.  However, their tendency to
superconduct can clearly be seen in the superconducting susceptibility
$\chi_{sc}(T)$;  for instance, in a quasi-one dimensional array of ladders,
one
can
estimate the superconducting $T_c$ from the mean-field equation $z{\cal
J}\chi_{sc}(T_c)=1$, where
$z$ is the number of nearest-neighbor ladders, and ${\cal J}$ is the
interladder
Josephson coupling.  For a
one-dimensional system with a spin-gap, the superconducting
susceptibility\cite{ekz,kfe}
is
\begin{equation}
\chi_{sc}\sim \Delta_s T^{1/K_c-2}
\end{equation}
where $K_c$ is the charge Luttinger exponent.  This expression reveals a
direct
relation between spin-gap  and enhanced superconducting fluctuations!
A related result, which can be more easily compared to the results of
numerical
experiments, is  the pair-field pair-field correlator, $D(r)$, given by
\begin{equation}
D(r) \sim \Delta_s |r|^{-1/K_c}.
\label{eq:Dofx}
\end{equation}

For the purpose of understanding the competition  between order parameters,
discussed below, it is worth
noting that the susceptibility to 2$k_F$ charge-density wave (CDW)  is
\begin{equation}
\chi_{cdw}\sim \Delta_s T^{K_c-2}.
\label{eq:CDW}
\end{equation}
A spin-gap enhances both the CDW
and  superconducting susceptibilities.  Indeed, it is tempting, and
sometimes
useful,
to think of the spin-gap as a superconducting gap, which exists in the
absence
of
superconducting long-range order due to  violent phase fluctuations
characteristic of
a one dimensional system. But  Eq. (\ref{eq:CDW}) makes it clear that the
spin gap can be equally well  thought of as a CDW gap, where true CDW order
has
been
suppressed due to the divergent fluctuations of the acoustic phonons in the
CDW
state.

The
charge Luttinger exponent, $K_c$,  is a non-universal function of doping,
$x$,
and
depends on the nature of the interactions.  As long as $K_c>1/2$, the
$\chi_{sc}$ is divergent, but not as
strongly divergent as the $\chi_{cdw}$, unless $K_c >1$.  For a Hubbard
chain,
it is well known that
$K_c < 1$, but this is  not a strict physical bound. For instance, by
comparing the expression
in Eq. (\ref{eq:Dofx}) with numerical experiments on the 2-leg Hubbard
ladder\cite{noack},  one can see  that
$K_c>1/2$ for a very wide range of parameters, and that for some ranges of
parameters,
$K_c >1$.  For instance, for  $x=0.0625$,  $K_c >1$ for
$5 \le U/t \le 15$, that is, for intermediate values of $U$.

\subsection{Hubbard molecules}

In the past we have examined a number of small Hubbard
molecules\cite{hmolecule} by exact
diagonalization and
found
pair-binding to operate in all cases. These were the $4\times 4$ torus, the
8-site
cube, and the 12-site
truncated tetrahedron. In all cases, the degeneracy of the state in which
doping
takes place
is important, and,   in all cases studied, second order perturbation theory
captures
qualitatively
all the aspects of the phenomenon for what appears to be large
values of $U$. This is not so surprising  in a finite system with
finite energy denominators;
hence,
perturbation theory is an effective tool in studying large molecules such as
C$_{60}$, where
exact
diagonalization methods are prohibitive.

We have also demonstrated that nearest neighbor repulsive
interactions for the truncated tetrahedron is harmful for pair-binding.
In contrast,
pair-binding in $t-J$ ladders remains robust\cite{dagotto} for nearest
neighbor repulsion, $V$,
as large as
$4J$.
Similarly,  realistic
frequency dependent screened potential for doped C$_{60}$ leaves the results
essentially
unchanged
from the pure Hubbard model\cite{Lammert}. This we could only demonstrate in second order
perturbation
theory.

The conclusion to draw is that in spite of the complexities at the
mesoscopic
scale, which surely
exist, it is well established that
certain mesoscopic clusters have positive pair-binding energies,
$E_p>0$.  Moreover,
$E_{p}$  can be of order the spin-gap, which is an intrinsically large
electronic energy scale.
There still remains the question  whether this mesocsopic pairing tendency
leads to global
superconductivity.

\subsection{The Two-Dimensional $t-J$ model}

Superconductivity in the two dimensional $t-J$ or Hubbard models
on a square lattice
remains controversial - no exact analytic results exist and
numerical studies are  confined to rather small systems or high
temperatures.  However, a few things are clear.
The undoped  system is an ordered antiferromagnet, and therefore
$\Delta_{s}=0$.  Moreover, while the pair-binding energy for two doped holes
is
positive on finite size molecules, in the physically relevant limit of
small $t/J <0.4$, $E_{p}$ decreases with system size, and
appears\cite{bonsegni} to
approach 0 in the thermodynamic limit.

It is plausible, even likely, that on a
sufficiently frustrated lattice, the spin 1/2 Heisenberg antiferromagnet
(the
``undoped system'') can have a quantum disordered ground-state with a
spin-gap\cite{sondhi}.  Such a system, when doped, might also become a high
temperature superconductor.  However, whereas pairing on the
mesoscale appears to be a very robust phenomenon, finding magnetic
systems with sufficient
frustration to produce a significant spin gap is not simple
in dimension greater than 1.  So while doping such a homogeneously
frustrated antiferromagnet
may provide a mechanism for high temperature superconductivity, we
find pairing on mesoscale structures to be a more common (and
hence more natural) mechanism.

\section{Effective interactions, competing orders, and superconductivity}

Since a finite cluster cannot be a superconductor, and even an
extended ladder cannot exhibit a finite temperature superconducting
state, the understanding of this mesoscale physics
simply serves to define a new, simpler effective Hamiltonian, which
governs the physics on longer length scales.

\subsection{Long-range order in molecular crystals}

When the extended system can be thought of as a molecular crystal
composed of finite clusters, the effective Hamiltonian is obtained
by integrating out all the molecular orbitals other than those at the
Fermi level.  Where there are several degenerate orbitals at the
Fermi energy, the resulting effective Hamiltonian will have more than one
molecular orbital per site.  To be concrete, we will use as an example the
case of a C$_{60}$-like crystal doped with $x$ added electrons per molecule,
with $0<x<6$ and  vary
the intra-molecular and the intermolecular couplings
to examine its properties.

The effective Hamiltonian is thus of
the form
\begin{equation}
H=\sum_{j} V({\bf S}_j^2,{\bf L}_j^2,\Delta N_{j})
+ \sum_{<ij>}H_{ij}
\end{equation}
where the effective interaction $V$ on the j$^{th}$  molecule depends on
 the total
number of added electrons $\Delta N_{j}$,
the square of the total spin ${\bf
S}_j^2$, and the square of the total orbital angular momentum, ${\bf L}_j^2$
(or more
precisely,  the equivalent Casimirs of the ichosohedral group),   and
$H_{ij}$ is the
intermolecular  interaction.  In the simplest case, $H_{ij}$ can be taken to
consist
of a sum of intermolecular hopping operators, which transfer a single
electron between molecules $i$ and $j$, but it can also contain
additional terms, such as assisted hopping and Coulomb interactions.

Where $E_{p}$ is positive on the cluster,
$V$ can be thought of as an effective attraction between pairs of
electrons on a given molecule.
The ratio of $E_{p}$ to the intermolecular band-width, $W$,
is an important parameter in the problem.

\subsubsection{Strong-coupling limit}

If the typical magnitude of $V$ is large compared to $W$,
we can further
thin the low-energy Hilbert space by integrating out all states other
than the molecular ground-states in each charge sector.
In most cases,
this limit is unphysical, but because it is easily analyzed, it is
worth considering for pedagogic purposes.  In
particular, if $E_{p}>0$, we can eliminate all states with an odd
number of electrons per site, while for $E_{p}<0$, we can
eliminate all states  with an even number of electrons.  The
resulting effective Hamiltonian has no remaining Fermionic degrees of
freedom;  the states are specified entirely by the number of
electrons, and possibly the spin, and the orbital moment of each
molecule.  Under appropriate circumstances, we are led to spin or orbital
effective Heisenberg
models, and even  to a bosonic version of the $t-J$ model.

Depending on details of the
interactions, such  effective Hamiltonians
can easily lead to a rich variety of ordered
phases, including ferro or antiferromagnets, orbital ferromagnets or
antiferromagnets, spin or orbital\cite{assa} nematics,
modulated (incommensurate or high-order commensurate)
versions of any of these states coexisting with charge-density wave
order, etc.  This is, of course, both exciting and discouraging.  It
suggests that in the presence of strong interactions in systems with
mesoscale structure, there should appear an enormous variety of
ordered phases, some of them  are new states of matter that have never
before been
documented.  While it invites us to exercise our imagination, the prospects
of predicting the
ultimate form  of order from microscopic considerations in any such system
becomes a difficult
task, if not impossible.

To take the simplest case, let us imagine that the effective
interactions favor an even number of electrons per molecule
($E_{P}>0$), and disfavor molecular states with non-zero spin or
orbital moment.  In this case, the low energy physics of the
molecular crystal reduces to that of interacting bosons (electron
pairs) on the molecular lattice
\begin{eqnarray}
H^{eff}=&& \sum_j [\mu_b b^{\dagger}_jb_j +V(b^{\dagger}_jb_j)]  \nonumber
\\
&& -  \sum_{<ij>}t_{\rm pair}[b^{\dagger}_jb_i+b^{\dagger}_ib_j +\ldots].
\end{eqnarray}
Here, $b^{\dagger}_j$ is a charge $2e$ bosonic creation operator,
$\mu_b$ is $1/2$ the electron chemical potential,
$V$ is the repulsion
between pairs on the same molecule and includes a hard-core interaction
that forbids more than three bosons on the same molecule,
$t_{\rm pair}$ is the pair hopping term which is second order in the
electron hopping
matrix element, and $\dots$ signifies additional interactions between bosons
on neighboring molecules.  Of
course, even here, the ground-state phases of the system will depend
on the precise balance between these residual interactions, as well as
on the mean boson density.

Typically, if the electronic
charge per molecule is $x=2$ or 4, the ground-state will be
insulating, with essentially a fixed number of electron pairs on each
molecule.  This state has no broken symmetry, but is a Mott
insulator, in the sense that in the non-interacting limit the system would
be
metallic.  For other mean-charge densities, this system will either
form a pair-density wave (presumably insulating) or a superfluid.  In
the latter case, the pairing scale is intramolecular, and hence large,
while the superfluid density, and hence the global phase ordering
temperature, $T_{c}$, will be smaller, determined by the
intermolecular bandwidth.  Moreover, in this case, pairing should
produce a pseudo-gap which survives at temperatures well above $T_{c}$.  In
this
limit, because the effective boson kinetic energy is proportional to the
square of the inter-molecular hopping matrix elements, one would expect
$T_c$ to be a decreasing function of lattice parameter, or an increasing
function of pressure.

\subsubsection{Weak coupling limit}

If the typical scale of interactions, $V$, is small compared to
$W$, and if the molecular crystal is three dimensional, it is
reasonable to treat the effective Hamiltonian in a
mean-field approximation, that is, by  variational ground-states, in the
spirit
of the BCS state, by solving the non-interacting problem in the presence
of a symmetry breaking mean-field.   Because of the narrowness of the bands
in typical molecular solids, it is equally unclear whether this limit is
actually
realized in real materials. But it is at least another limit  in
which controlled theoretical results can be obtained.  It is probably closer
to reality than the strong coupling limit, at least
in relatively metallic materials, such as alkali doped C$_{60}$.

In this
limit, the various charge, spin, and orbital moment density wave states are
unlikely, unless there
are particular nesting wave-vectors of the relevant Fermi surface.  Thus,
except at certain
commensurate values of $x$, they can be safely ignored.

For
positive $E_p$, a BCS instability to a superconducting state is still
generic with a
superconducting transition temperature,
\begin{equation}
T_c\sim W \Theta(V^{\rm eff})\exp[-1/\rho(E_F)V^{\rm eff}(x)]
\label{eq:Tc}
\end{equation}
where $\Theta$ is the Heaviside function, and
\begin{equation}
\rho(E_F)\propto 1/W
\end{equation}
is the density of states at the Fermi energy of the non-interacting
band-structure.
The quantity
\begin{equation}
V^{\rm eff}(x)= \sum_{Q,S} P_x(Q,S) E_p(Q,S)
\end{equation}
is the average of the pair-binding energy for the various charge
states of the molecule
averaged over the appropriate probability distribution
($\sum_{Q,S} P_x(Q,S)=1$) of charge
and spin-states of the molecule.  Here, $T_c$ would be expected to be a
strongly
increasing function of lattice parameter (decreasing function of pressure)
since
it varies with the inverse exponential of $1/W$.

While it might seem at first that this  BCS expression for $T_c$ is generic
to
any mechanism of superconductivity with an
intra-molecular pairing force (for instance,
one generated by coupling to the intra-molecular optical phonons), the
strong $x$
dependence of $V^{\rm eff}$ implied by the  electronic mechanism discussed
here
is very pronounced and unique.
For instance, from the perturbative results for the Hubbard model on a
C$_{60}$
cluster
described above\cite{cgk}, one might expect that for
intermediate $U$, $V^{\rm eff}$ is positive (attractive) for $x=3$,
and negative (repulsive)
for $x=2$ or $x=4$.  Crudely, this leads to a doping dependent effective
attraction of the general form
\begin{equation}
V^{\rm eff}= V_0  (x-x_0)(x_1-x),
\label{eq:V0}
\end{equation}
with $2 < x_0 < 3 < x_1 < 4$. This leads to a large $T_c$
only in a constrained region about
$x=3$, and to $T_c=0$ for $x<x_0$ or $x>x_1$.  This is illustrated in Fig.
5,
where, to make a crude comparison with experiments, we have
taken $W=0.1$eV, $\rho(E_F)V_0=1/3$, $x_0=2$, and $x_1=4$.

\begin{figure}[htb]
\centerline{\epsfxsize=3 in\epsffile{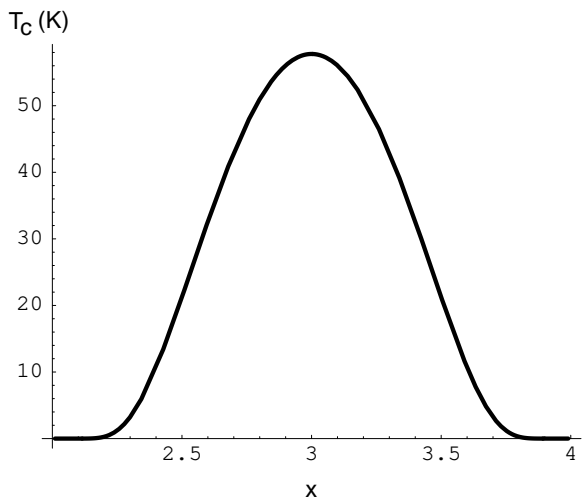}}
\caption{\protect $T_c$ as a function of electron concentration $x$,
for the simple, weak-coupling model of correlation-induced
pairing in C$_{60}$ described in Eqs. (\ref{eq:Tc}) and (\ref{eq:V0}).}
\label{fig5}
\end{figure}

\subsection{Long-range order in arrays of ladders}

Long-range order in an array of ladders is related to
the general problem of order in quasi-one dimensional electronic
systems and has been studied for for many years\cite{bourbonais};  this is
theoretically one of the best
understood cases of competing orders.  In particular,  the occurrence of
a spin gap
enhances the tendency for both CDW and superconducting  order
in a way that does not
distinguish between them.

Even in this case, the competition plays out in a manner that was not
previously recognized.  It is not sufficient to simply identify the most
divergent
susceptibility of the isolated ladder in order
to determine which type of long-range order will win out.  There are,
in addition, a variety of marginal interladder forward scattering
interactions which can affect the balance, or whether or not there is an
ordered ground-state at
all\cite{efkl}.  A cautious, but optimistic conclusion
from the mesoscale calculations is that the spin-gap, which can be
large on narrow ladders, permits the {\it possibility} of high
temperature superconductivity, but  it can also lead to a variety of
alternative insulating or  metallic phases.

\section{Final remarks}

Superconductivity in simple metals is well understood in terms of the
BCS mechanism --- the normal state is well approximated as a Fermi
liquid, and the effective attraction between electrons is mediated by
phonons.  The effects of electron-electron interactions appear in
this description in the form of a few Fermi liquid parameters (which
are typically small), and in the Coulomb pseudo-potential, $\mu^*$.  The
Fermi liquid character
of simple metals is exceedingly robust - in the absence of special nesting
conditions of the
Fermi surface or very strong interactions, the {\it only} instability
of the metallic  state is the BCS instability.
A conventional superconductor, when driven normal by suitable
alloying, by weak disorder
(magnetic impurities), increasing temperature, or magnetic field, remains
a highly conducting metal.

In both the cuprates and the organic superconductors,
including doped C$_{60}$, there is strong reason to doubt the
accuracy of the Fermi liquid description of
the normal state. Screening is generally poor, and the typical
magnitude of Coulomb energies is large compared to the valence-electron
bandwidth.  And, indeed, the materials are not robustly metallic -
resistances greater than the quantum of resistance (often accompanied
by ``insulating'' negative values of $d\rho/dT$) are frequently found in these
materials when the superconducting state is destroyed.  
Even when the resistance has a metallic temperature
dependence (positive $d\rho/dT$), as it does in doped  C$_{60}$
in the ``normal'' state, the resistance is too large to correspond to
any sort of freely propagating quasi-particle.  In electron-doped C$_{60}$,
the room temperature resistance\cite{hebbard,bertram2} is 2-5 m$\Omega$-cm, 
which were
it interpreted in terms of a Drude conductivity would correspond to a
mean-free path of 1-2 {\AA};  in  hole doped 
C$_{60}$, the room temperature resistance\cite{bertram2} is almost 20m$\Omega$-cm,
which at face value would correspond to a mean-free path of order 0.1{\AA}!  Finally,
in addition to superconducting order, evidence of a variety of competing
orders abounds.

Given the striking  evidence of
strong correlation effects in these
materials, we suppose that they are
fundamental to the mechanism of their unprecedentedly high $T_{c}$'s.
However, we suggest that such a mechanism is robust
only if the pairing originates on mesoscale structures.  Therefore,
it is not accidental that the materials in question are either
molecular crystals, with reasonably large molecular building-blocks,
or materials which exhibit  self-organized mesoscale
structures.

\subsection{The electron-phonon coupling}

Electron-phonon interactions are reasonably strong in all the
high temperature superconductors.  In particular, the
phonon frequencies in C$_{60}$ are much higher than in
conventional metals.  In spite of this, we very much doubt
that superconductivity is driven principally by electron-phonon
interactions.
That $T_c$ exhibits isotope effect  cannot be an argument
as it can be explained by the electronic mechanism as well\cite{isotope}.

In the first place, the pairing induced by strong repulsive
interactions on all the clusters that have been studied to date
have  d-wave character, for reasons that are by now well
understood.  Phonons typically produce $s$-wave attraction, and
are often pair-breaking in other channels.  Indeed, when
we\cite{salkola}
analyzed the effects of electron-phonon couplings on an assumed
electronic mechanism of pairing on a C$_{60}$ molecule, we found
that most, but not all of the high energy phonons are pair-breaking.
Note that local $d$-wave character may or may not imply
that the global order has d-wave symmetry --- for instance, we have
found\cite{ck,cgk} that the superconducting order in electron-doped C$_{60}$
is globally
s-wave.

In the second place, the strong electron-phonon interaction,
presumably responsible for high $T_c$, is always accompanied by
self-trapping, or bipolaron formation.  The
result  is  a large exponential Frank-Condon  reduction of the energy scale
for coherent motion of
charge, inevitably leading to an insulating state.

In the
weak-coupling limit, the usual BCS result applies with an
exponentially small pairing scale,  and the retarded nature of the
electron-phonon coupling
plays an essential role --- the Coulomb interaction between electrons
is reduced due to the fact that they do not have to be at the same
place at the same time.  However, in the cuprates, and even more
clearly in doped C$_{60}$, the phonon frequencies are comparable to
the electron bandwidth, so retardation does not lead\cite{khlebnikov} to
any reduction of $\mu^{*}$.

\subsection{Superconductivity in the polyacenes}

Superconductivity has been recently discovered in a surface layer of
 electrostatically
doped crystals of anthracene, tetracene, and pentacene\cite{bertram3}.
These
molecules can be viewed as segments of a two-leg ladder with every other
rung
weakened.  Thus, we believe that the basic mechanism of
superconducting pairing is the same as the one discussed previously  for the
two leg $t-J$ and Hubbard ladders.  As a consequence, we are lead
to expect that the superconducting order will have a $d$-wave-like symmetry,
the order parameter changing sign under $90^{o}$
rotation.   However, because of the quasi-one dimensional geometry, it is
possible
that there are no gapless ``nodal'' quasi-particles in this system,
despite this\cite{granath}.

We note that a number of features of the experiment are consistent
with this suggestion.  In the first place, $T_{c}$ is maximal when
there is approximately an odd number of electrons per molecule.   Moreover, the fact that
the optimal $T_{c}$ decreases with the increasing size of the molecule is
consistent with a mesoscopic origin of the pairing.

\subsection{C$_{60}$}

Narrow bands in C$_{60}$ make superconductivity at high temperatures (presently 52 K) difficult
unless the molecule itself is in some sense superconducting. The same difficulty holds for all
narrow band materials\cite{pwa2} for the following reasons. 

First, the net interaction at
$\omega=0$ must be repulsive, otherwise we will get a CDW, that is, the interaction is 
\begin{equation}
V_{\rm eff}(q,\omega)= \frac{4\pi e^2}{q^2\epsilon(q,\omega)}, \  \epsilon(q,\omega=0)>0 .    
\end{equation}
Typically, a peaked density of states implies that only a single band is relevant, corresponding
precisely to the band in which the peak in the density of states is situated. Thus,   if only
one band is relevant, no local field effects are possible, and the ions will see the same
$\epsilon$ as the electrons. 

The second reason is that the retardation of the electron-phonon
interaction in BCS superconductors 
(that is so crucial for the existence of a net electron-electron
attraction through the reduction of the Coulomb pseudopotential) is no longer  operative. This is
because the Coulomb pseudopoetntial $\mu^*$ is given by
\begin{equation}
\mu^*=\frac{\mu}{1+\mu\ln \frac{W}{\omega_0}}\ ,
\end{equation}
where $\mu$ is the Coulomb matrix element at the bare high energy scale,  $W$ is the band width,
and $\omega_0$ is the typical phonon scale. But if
 $W$ is not much larger than $\omega_0$, which is manifestly the case in C$_{60}$, we get little
reduction in
$\mu^*$. 

In contrast, in a molecular crystal, the existence of a structure at an intermediate scale implies atomic energy scales much
larger than intramolecular energy scales, which in turn are larger than the intermolecular scales of the solid. This clear
separation of scales allows us to solve first  the problem of a single C$_{60}$ molecule alone and then incorporate the
coupling between the molecules. This approach leads to the 
%unusual, but physically relevant 
by now well substantiated phenomenon of attraction from
repulsion that has been the corner stone of our approach. 
%Even the initial issues regarding the precise modeling of the
%molecule is now fully resolved
Even issues concerning the proper screening of the longer range
pieces of the Coulomb interaction can be reliably analyzed\cite{Lammert}
taking advantage of this hierarchy of energy scales.

We conclude with a few observations regarding the
present experiments\cite{bertram1,bertram2} on C$_{60}$ FET's. 
It is clear that high density
of
states cannot be the principal feature responsible for the high
temperature superconductivity in hole doped C$_{60}$
as $T_c$ is clearly not peaked at the same charge per molecule as the density
of states.
Moreover, the $T_c$ profile as a function of doping
does not presently have
any natural explanation in the electron-phonon
theory of a BCS superconductor;  there appears to be no reason for the
electron-phonon
coupling to show such a strong dependence on the band 
filling. In the electron doped case\cite{bertram1}, the onset of
superconductivity
is very sharp. This is consistent with  our theory in which $T_c$ is
highest at
a doping corresponding to an odd number of added electrons or holes,
where the pair-binding on a single molecule
is largest, but then vanishes for doping corresponding to an even number of added 
electrons or holes\cite{cgk}.  A crude model
was
shown in Fig.\ref{fig5}. Of course, sample inhomogeneity and disorder can
broaden the doping profile, which one should be able to
better address in the near future as the sample quality improves further.
Moreover, relatively {\it low temperature tails} in $T_{c}$ vs $x$ could
result from residual electron-phonon interactions.

If the explanation of superconductivity lies 
in the electronic correlation effects on the mesoscopic scale, there 
is no reason why $T_c$ could not be raised further ($>52$ K) in the near future.
Such high values of $T_c$ will constitute a strong argument against the electron-phonon
mechanism of superconductivity in these materials.

\acknowledgments

S. C. is supported by NSF-DMR-9971138.
The work was also conducted under
the auspices of the DOE, supported by funds
provided by the University of California for the conduct
of discretionary research by Los Alamos National Laboratory.
SAK was supported, in part,
by NSF-DMR98-08685 and DE-FG03-00ER45798 at UCLA.

\appendix
\section{Perturbation theory for Hubbard chain}
\label{sec:append}
First order perturbation theory
leads to the conclusion, which turns out to be exact\cite{shiba} for all
$U/t$,
that for $N=4n$, the ground-state is a spin
singlet for $Q=0$ and $Q=\pm 2$ and a doublet for $Q=\pm 1$, while for
$N=4n+2$ the ground-state is a singlet for $Q=0$, a doublet for
$Q=\pm 1$, and a triplet for $Q=\pm 2$.  As a consequence,
\begin{equation}
E_{p}(0)=\frac{A_{2}(N)}{N}\frac{U^2}{t} + {\cal O}(U^{4}/t^{3})
\end{equation}
where $A_{2}$, a number of order unity, is positive for $N=4n$
and negative for $N=4n+2$, reflecting the difference in
spin of the doubly charged state. 

For $N=4n+2$  and  $U=0$, the ground-state with
$Q=0$ is a unique
spin singlet with a gap;  for $Q=\pm 1$,
the ground-state
is still unique up to symmetry --- it has spin 1/2 and non-zero crystal momentum;
however, for $Q=\pm 2$, there are three singlet and one spin triplet
states that are 
degenerate.  In first order in $U/t$, the degeneracy of the
doubly charged system is lifted and, as required by Hund's rule, the
spin-1
state is selected as the ground-state.  In the Hubbard
model, there is no direct interaction between electrons of like spin, so a
consequence of the triplet character of the doubly charged state is
that there is
no first order contribution to the pair-binding energy.
However, mainly because of the unfavorable ferromagnetic
correlations in the
doubly charged state, it turns out that $A_2(4n+2)$ is negative ---
there is pair repulsion.

For $N=4n$, there are two degenerate orbitals at the Fermi level of the
non-interacting  neutral molecule resulting in accidental degeneracies.  To first order in
$U/t$,
the energies of two of the singlet states of the neutral molecule are
elevated,
but one singlet state remains degenerate with the triplet state.  This
miracle is due to an Umklapp process which allows a singlet pair of electrons,
each with
crystal momentum $\pi/2$, to scatter into the state in which each
electron has crystal
momentum $-\pi/2$.  The final degeneracy for $Q=0$ is then lifted in
second order
perturbation theory, in which the antiferromagnetic correlations of the
singlet state
are preferred over the ferromagnetic (triplet) state.  This violation of
Hund's rule is
peculiar to the case of $N=4n$ and $Q=0$.  By the same token, the
antiferromagnetic correlations of the neutral molecule lead to anomalous
stability of
the neutral state, and hence a negative $A_2(4n)$ --- there is pair
binding.  It is worth
noting that, although for large $U/t$ the perturbative results are, of
course,
quantitatively poor, as there are\cite{shiba} no groundstate level
crossings
as a function of
$U/t$, perturbation theory gives correct results for the ground-state
quantum numbers of
all these states.

\end{document}